# Negative Capacitance and Inverted Hysteresis: Matching Features in Perovskite Solar Cells

**Agustin O. Alvarez[1]; Ramón Arcas[1]; Clara A. Aranda[2,3*]; Loengrid Bethencourt[1,4]; Elena Mas-Marzá[1]; Michael Saliba[2,3]; Francisco Fabregat-Santiago[1*]**

[1]*Institute of Advanced Materials (INAM), Universitat Jaume I, 12006 Castelló, Spain*

[2]*IEK-5 Photovoltaics, Forschungzentrum Jülich, 52425 Jülich, Germany*

[3]*Institute für Photovoltaik (IPV), Universität Stuttgart, 70569 Stuttgart, Germany*

[4]*Grupo de Desarrollo de Materiales y Estudios Ambientales, Departamento de Desarrollo Tecnológico, CURE, Universidad de la República, Ruta 9 Km 207, Rocha, Uruguay*

*Email:* fabresan@uji.es; caranda@uji.es

## Abstract

Negative capacitance at the low-frequency domain and inverted hysteresis are familiar features in perovskite solar cells, where the origin is still under discussion. Here we use Impedance Spectroscopy to analyse these responses in methylammonium lead bromide cells treated with lithium cation at the electron selective layer/perovskite interface and in iodide devices exposed to different relative humidity conditions. Employing the *Surface Polarization Model*, we obtain a time constant associated to the kinetics of the interaction of ions/vacancies with the surface, $\tau_{kin}$, in the range of $10^0$ - $10^2$ s for all the cases exhibiting both features. These interactions lead to a decrease in the overall recombination resistance, modifying the low-frequency perovskite response and yielding to a flattening of the cyclic voltammetry. As consequence of these results we find that that negative capacitance and inverted hysteresis lead to a decrease in the fill factor and photovoltage values.

**Table of Contents (TOC)**

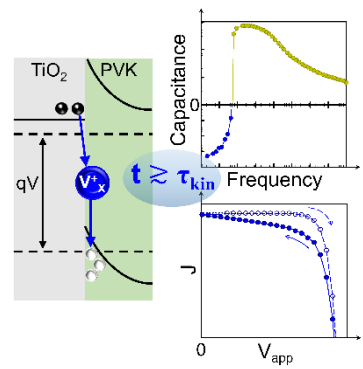

The progress of lead halide perovskite solar cells (PSCs) in the photovoltaic field has been astoundingly rapid.[1] However, the current impressive power conversion efficiency (PCE) approaching 26% in the lab scale, is accompanied by a wide range of features during device operation that still needs to be solved.[2] Hysteretic behaviour is one of them, representing a bottle-neck in the reporting of PSCs performance for high-efficiency values, slowing down the overall development.[3,4,5] This feature consists of a delay between the change of the system properties under variation of an external electric field. The consequence is a difference in the J-V curves during the sweeping in the two directions from short-circuit towards open-circuit (Forward scan, FS) and *vice versa* (Reverse scan, RS). A noteworthy manifestation of this phenomenon depends on the external parameters affecting the J-V curve: i.e. temperature, light intensity, scan-rate, pre-biasing voltage and external contacts. These perturbations lead to a wide range of hysteresis responses including: photocurrent decay in RS, apparent $V_{oc}$ shift in FS, capacitive hysteresis and *bump* (maximum current close to $V_{oc}$ at the FS).[6,7,8] The hysteretic behaviour most commonly observed in PSCs is the *normal hysteresis* (NH) which performs with an improved fill factor (FF) and photovoltage ($V_{oc}$) for the RS, while the FS brings lower values. In this process, the cell behaves as a capacitor. During the FS, an accumulation of charge occurs ("capacitor filling"), while in the RS this charge is realised ("capacitor emptying") leading to an excess of current in the external circuit. On the other hand, an *inverted hysteresis* (IH) can also be present in certain samples and/or under specific preconditions. In this case, the RS displays lower photovoltage and FF values than the FS. Several mechanisms have been proposed for the different hysteretic features, including charge trapping and de-trapping, ferroelectric polarization, ion migration, capacitive effects, charge accumulation at the interfaces and unbalanced distribution of electrons and holes.[9,10,11,12] All of them are connected with the mixed ionic-electronic nature of PSCs, bringing out the perovskite/selective contact interfaces as the main character controlling charge extraction capability, accumulation and recombination mechanisms.[13] Additionally, the accumulation mechanisms reported for the IH involve unfavourable energy level alignment, ionic accumulation and space-charge buildup.[14,15]

Tress and co-workers observed a NH for simple $MAPbI_3$ while an IH for mixed-cation perovskite devices. After including an isolating layer onto the mesoporous titanium dioxide (m-$TiO_2$) in the simple perovskite device, they concluded that the process responsible for the IH was an energetic extraction barrier at the $TiO_2$ interface.[16] On the other hand, Nemnes et. al showed that pre-poling bias (in particular negative pre-poling) can produce the switching from NH to IH for the same structure device. They also analysed by a *dynamic electrical model* (DEM) the influence of the scan rate in terms of the time evolution of electronic and ionic charge accumulated at the interfaces. This DEM model includes a capacitor accounting for nonlinear polarization effects in the equivalent circuit.[17] This theory is in perfect agreement with the *Surface Polarization Model* (SPM) proposed by Bisquert and coworkers.[13] This model is based on the formation of a surface potential, $V_s$, at the electron selective layer (ESL)/perovskite interface due to the arrival of positively charged ions under large injection or illumination conditions, creating an accumulated density of both ionic and electronic holes, increasing recombination dynamics. An experimental work reported by Bisquert and co-workers provided evidence of the surface polarization of the ESL/perovskite interface.[15] In their work, a switch from NH to IH is shown when the thickness of the c-TiO$_2$ material is diminished, attributing this evolution to an interaction between charge accumulation and recombination mechanisms. The charge accumulation at the TiO$_2$/perovskite interface has been related with an atypically high capacitive feature (mF cm$^{-2}$) in the low-frequency domain (LF) ($<$ 1 Hz), measured using Impedance Spectroscopy (IS) and being also present under dark conditions, which brings out the strong influence of slow ionic dynamic on the hysteretic behaviour.[12] This relation was demonstrated in the work of Park and co-workers.[18] They systematically compared the hysteretic behaviour between a regular perovskite configuration, using TiO$_2$ as ESL and an inverted configuration with PCBM as ESL. In the case of inverted structure, a hysteresis-free performance of the J-V curve was observed together with a much lower value of capacitive response in the LF domain. They established the origin of this reduction as a decreased polarization of the device electrode.

Beyond the giant capacitive response, even in dark conditions, a negative capacitance has also been measured under a wide range of external conditions and in many different perovskite devices, becoming one of the hardest features to decode.[19,20,21,22] The only thing in which there seems to be some agreement in literature is the unfavourable effect on the optoelectronic response of the cells.[23,24] An electron injection through interfacial states was attributed to the origin of negative capacitance in light-emitting diodes (LEDs), assigned as well as the main cause in PSCs by Anaya et. al.[25] Bisquert and co-workers analysed the IS response of samples with negative capacitance by SPM,[26] matching the negative capacitance to an inductive element included in their equivalent circuit (EC). This negative capacitance was attributed to a delay of the surface voltage depending on a certain kinetic relaxation time, controlling the equilibration of the ESL/perovskite interface governed by the ionic dynamic. The value of that kinetic relaxation time depends on the process governing the slow response. To this respect, a recent review summarizes the kinetic relaxation time values for halide vacancy diffusion and the response of ions accumulated at the perovskite surface; being in the range of 10$^{-1}$ s and 10$^0$ – 10$^2$ s, respectively.[27]

The next question lies then if there is any relation between the negative capacitance and hysteresis response. A recent work reported by Tress and co-workers correlates the negative capacitance with a stronger scan rate dependence of hysteresis behaviour, considering a change in the charge injection as the cause of this effect.[28] Contrary, Jacobs et.al reported an ion-drift-diffusion (IDD) theory to simulate the negative capacitive process and its dependence on scan rate, pointing to a phase-delayed recombination mechanism due to mobile ions.[29]

In this work we demonstrate experimentally how negative capacitance and inverted hysteresis are related and the common origin by which both effects are produced. We apply the SPM to analyse the impedance response of a batch of MAPbBr$_3$ devices with lithium treatment at the ESL[30] and MAPbI$_3$ devices exposed to moisture. Matching results are obtained for all the cells. Samples with IH and strong negative capacitive response present higher kinetic relaxation times and lower recombination resistance values than the ones with lower IH or NH and low/null negative capacitances. We correlate these results with a surface ionic interaction, leading to a delayed recombination mechanism. These insights provide a concise explanation about the origin of these two features, confirming the direct relationship between them and advancing forward to a deeper knowledge about PSCs dynamics.

We evaluate in the first place the hysteretic behaviour of two representative champion devices of well-known MAPbBr$_3$ solar cells with and without lithium treatment at the ESL (Li and No-Li samples, respectively).[30] As previously demonstrated, this Li$^+$ treatment leads to an increased photovoltage value, enhancing the electroluminescence (EL) emission of the perovskite bulk, due to a reduction of undesired non-radiative and surface recombination mechanisms by reducing the density of holes in the charge accumulation zone. The ionic nature of this effect may have a strong influence on the cyclic voltammetry (CV) measurement at different scan rates.[4,6] Then, CV from 10 mV s$^{-1}$ to 500 mV s$^{-1}$ under illumination conditions are developed in order to force the ionic effect into the photogenerated electronic carriers.

In Fig. 1a and 1b we show the corresponding hysteretic behaviour for champion cells with a without lithium at the ESL, respectively. The strong differences between them are clear determining the hysteresis index calculated from eq.1:[18]

$$HI = 1 - \frac{PCE_{FS}}{PCE_{RS}} \tag{1}$$

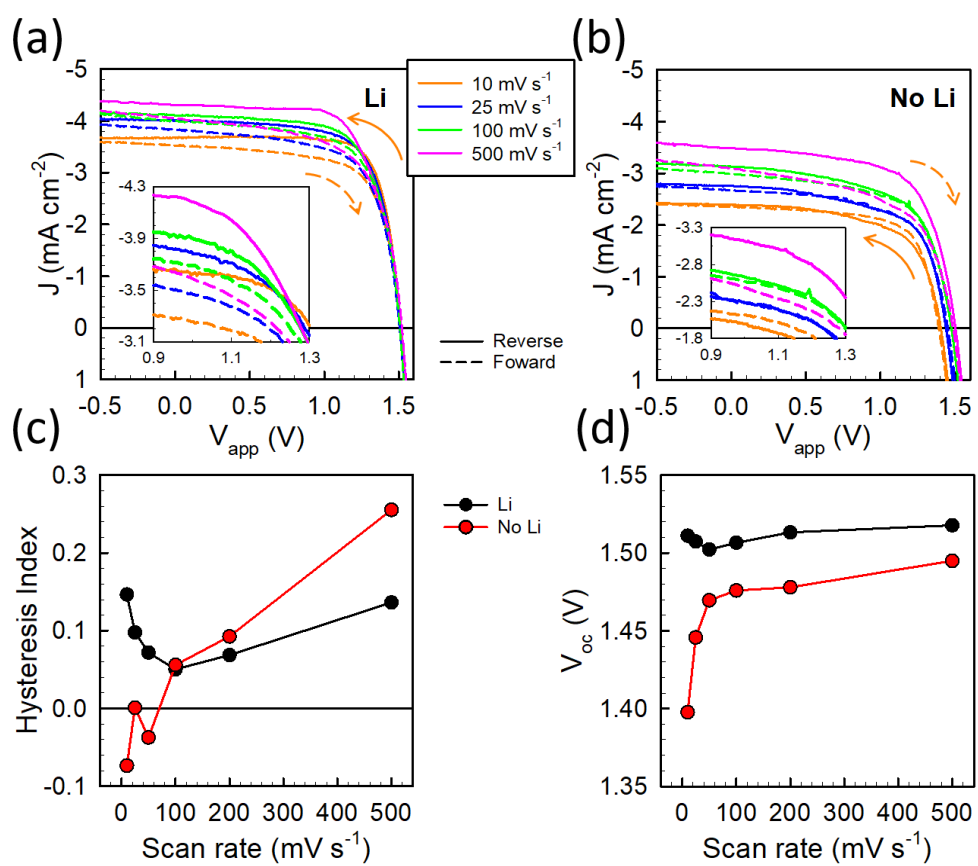

**Figure 1.** Cyclic voltammetry of Li (a) and No-Li MAPbBr₃ cells (b) devices performed at different scan rates from 10 mV s⁻¹ to 500 mV s⁻¹. (c) Hysteresis index and (d) photovoltage *versus* scan rate comparing both samples: black dots for doped and red dots for un-doped samples.

As it can be observed in Figs. 1a and 1c, the Li device presents NH independently of the scan rate applied (positive hysteresis index). However, the No-Li device (Figs 1b and 1c) exhibits a more pronounced NH at high scan rates (bigger hysteresis index) and IH at low scan rates (negative hysteresis index). The variations of the photovoltage with the scan rate are also notorious comparing both type of samples (Fig. 1d and Fig. S1 in SI). While the open circuit potential of the Li sample is almost constant with the scan rate, in the absence of lithium the photovoltage increases sharply for scan rates from 10 to 50 mV s⁻¹ and then presents little variations. Furthermore, the photovoltage value is lower for the No-Li device at any scan-rate. This is due to the major surface recombination, previously reported.[30] Comparing Figs. 1c and 1d, it can be noted that, in general terms, when the HI diminishes, so does the V$_{oc}$ (see Figure S1 in SI). Besides, for both cases, an improvement of photocurrent proportional to the scan rate is found (see Fig. 1 a, b and S1 in SI). This effect has been attributed previously to a slow process related to the build-up of the space charge close to the contacts, due to ionic displacement and associated with a capacitor-like discharge of current.[7,31]

The negative capacitance feature has been considered as a deleterious effect for the PSCs performance.[24,20] Then, to evaluate the nature of this dynamic, we performed an analysis of the impedance response and the hysteretic behaviour of samples presenting moderate PCE. In this context, we consider moderate PCE for devices with photovoltages below 1.5 V for Li samples and below 1.4 V for the No-Li devices. All the good cells (high PCE) are considered above these values respectively. We establish these limits taking into account the record photovoltage value of 1.56 V reported for MAPbBr$_3$ by Aranda et. al[30]. The cyclic voltammetry of the two devices (Li and No-Li) performed at 50 mV s$^{-1}$ are shown in Fig. 2a.

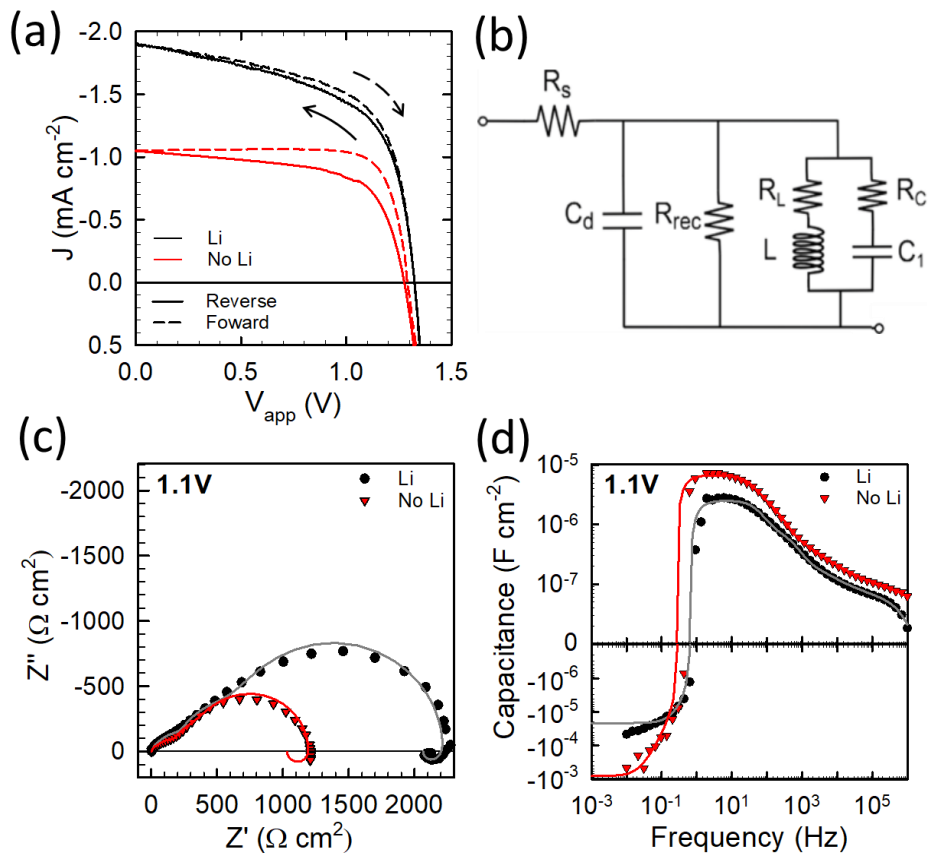

**Figure 2.** Black and red colours correspond to the Li and No-Li devices respectively. (a) CV measures showing inverted hysteresis for both configurations. The dashed lines correspond to forward scans, while the solid lines represent the reverse ones. (b) EC established by the SPM and used here to fit our experimental impedance results. R$_s$ is the series resistance, C$_d$ is de dielectric capacitance, R$_{rec}$ is the recombination resistance, R$_L$ is a resistance associated to the inductance L and R$_C$ is the series resistance associated to the interfacial charging capacitance C$_1$. (c) Impedance plot and corresponding (d) capacitance spectra, measured under dark conditions and at 1.1V. Solid lines correspond to fits using the equivalent circuit showed in (b).

As it can be observed, both samples present an inverted hysteretic behaviour, but clearly, this effect is much stronger in the case of the No-Li device. Capacitance and hysteretic phenomena have already been correlated with slow transients, specifically to ionic electrode polarization. Therefore, we performed the IS analysis at different potentials towards open-circuit and under dark conditions to prevent further complexity due to photogeneration mechanisms (see Fig.S4).[12] In Fig. 2c it is shown the fitting of the impedance spectra measured at 1.1 V (where the effect is more clear due to the greater polarization of the sample) and in a wide range of frequencies from 1 MHz to 10 mHz, to ensure the low-frequency ionic contribution. As it can be observed, pronounced differences in the recombination resistance are presented comparing both samples (Z' axis cross). This is in complete agreement with our previous results, confirming that lithium decreases the surface recombination dynamics. Together with that, a negative capacitance appears for both samples, being much more pronounced for the No-Li device, (Fig. 2d). This negative capacitance is represented as the inductance in the $R_L$-L line of the EC proposed by Bisquert and coworkers in their SPM (Fig.2b), used as well in other works[32]. This new line produces the low frequency arc below the real axis which is the clearest signature on impedance of this phenomena or, in other situations, a loop that makes two arcs in the impedance to cross each other.[33,34] The extracted parameters from the impedance spectra corresponding to both samples, using the EC of Fig.2b, are detailed in Table 1 (See Table S1 for the complete fitting parameters).

**Table 1.** Extracted parameters from dark IS measurements for the samples with and without Li$^+$ at the ESL.

| Cells | $R_{rec}$ ($\Omega\cdot cm^2$) | L (H$\cdot cm^2$) | $R_L$ ($\Omega\cdot cm^2$) | $\tau_{kin}$=L/$R_L$ (s) |
|---|---|---|---|---|
| Li | 2.2 x 10$^3$ | 2.1 x 10$^4$ | 2.9 x 10$^4$ | **0.7** |
| No Li | 1.2 x 10$^3$ | 5.6 x 10$^4$ | 9.9 x 10$^3$ | **5.7** |

The SPM establishes that the surface potential ($V_s$), created by the ionic accumulation at the ESL, cannot follow immediately the external voltage ($V$) variations.[13,26] The relaxation equation defining this behaviour was described by the SPM as follows:

$$\frac{dV_s}{dt} = -\frac{V_s-(V-V_{bi})}{\tau_{kin}} \qquad (2)$$

where $V_{bi}$ is the built-in potential (see SI for the definition of the EC elements corresponding to the model, equations S1-S5). Here $\tau_{kin}$ represents the characteristic kinetic relaxation time of this delay which is governed by the ionic movement speed and their possible accumulation to the surface. This induces an increase of the recombination processes and generates the negative capacitance (modulated by the inductance) (see SI for further explanation).

The recombination resistance value ($R_{rec}$) for the Li device almost doubles the No-Li device (Table 1), confirming a reduction of recombination losses due to lithium effect. According to the model, $R_L$ and $R_{rec}$ are both inversely proportional to $\bar{J}_{rec}$ (see eq. S2 and S3). In the EC they are in parallel, consequently, $R_L$ reduces the total recombination resistance at low frequency. The smaller $R_L$ is, the smaller total recombination resistance will be. In Table 1 can be noted that $R_L$ and $R_{rec}$ are smaller in the case of the No-Li sample, what provides the smaller low-frequency resistance limit in Fig. 2c. Additionally, the $\tau_{kin}$ value for the No-Li sample is one order of magnitude bigger than Li sample. Large values of the kinetic relaxation time are attributed to slower ionic dynamics and long delays to reach the equilibrium state after an external voltage is applied ($V$). For the No-Li device, the kinetic time value is in the order of $10^0$ - $10^2$ s, which agrees with a response of accumulated ions at the perovskite surface interacting with contacts (Ref. 26). In fact, the reactivity of migrating ions with $TiO_2$ defects ($Ti^{3+}$) have been discreetly reported in the literature as other cause of instability and hysteretic responses.[35,36,37] A large intrinsic accumulation of ions (or vacancies) at the ESL/perovskite interface could facilitate their interaction at the surface, yielding larger kinetic relaxation times and increasing surface recombination processes. On the other hand, the sample with Li at the ESL/perovskite interface shows a kinetic time value in the range of $10^{-1}$ s, which agrees with halide vacancy diffusion time, indicating "free" bulk ionic movement.[27] When ions/vacancies move freely, negative capacitance and inverted hysteresis are very limited (usually not observed). The presence of lithium at that interface could reduce the accumulation of these bulk ions, avoiding their interactions with the surface, as indicated by the small value of $\tau_{kin}$ obtained.

Besides, it has recently been reported the photoluminescence (PL) quenching as an effect of the ion drift-diffusion.[38] The origin of this phenomenon is attributed to the creation of non-radiative recombination centres in the perovskite bulk. To reduce this non-radiative recombination mechanism, potassium doping of the perovskite material has been used to avoid the ionic attachment to the surface.[39] Similar effect was found in our previous work, where lithium cation at the ESL generates an additional effect enhancing the PL and electroluminescence (EL) response of the $MAPbBr_3$ bulk material.[30] This phenomenon could explain the recombination decrease when associated with vacancy diffusion kinetics, obtained for the Li sample.

These results are not only a consequence of the specific conditions of bromide cells and we prove it performing the same analysis in the most widely used perovskite device: $MAPbI_3$. In these standard cells under dry working conditions, there is no presence of IH or negative capacitance, thus we induced it in a controlled way. Moisture is one of the main degradation factors in PSCs, closely related to hysteretic behaviours.[40] Thus, several IS and CVs measurements were performed on $MAPbI_3$ cells under different relative humidity (%RH) conditions (see Fig. 3 and the experimental part in SI).

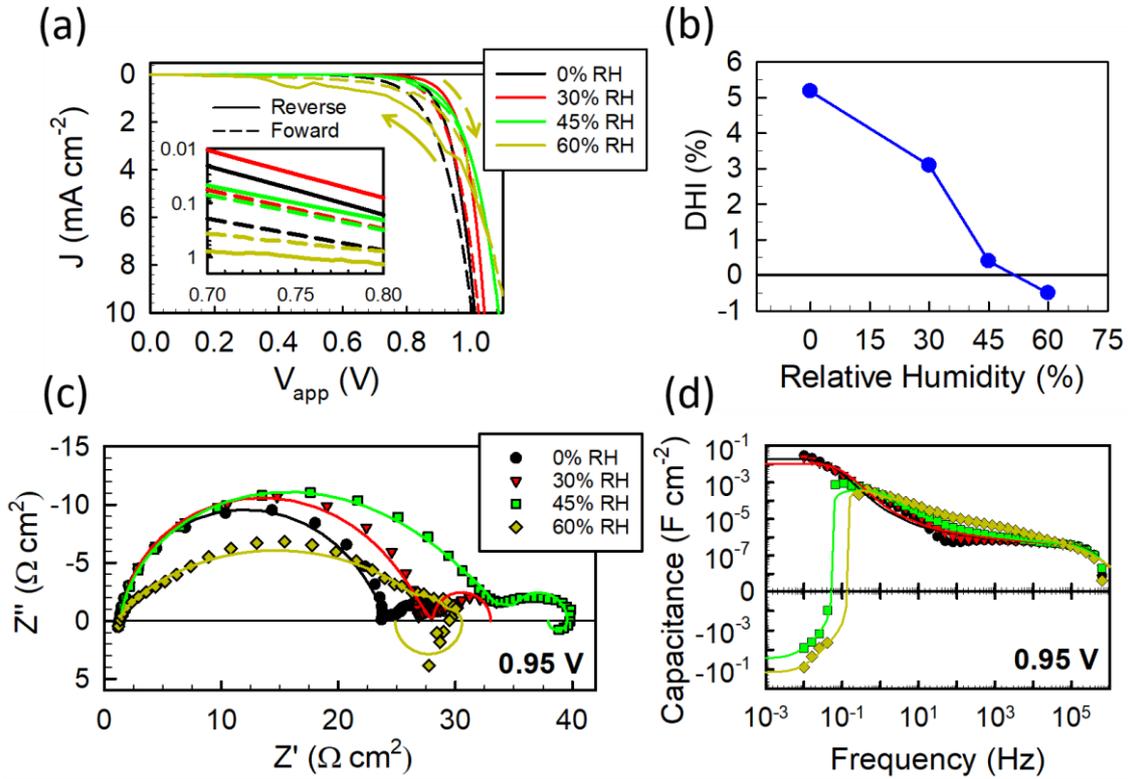

**Figure 3.** Measurements of regular MAPbI₃ solar cells (FTO/TiO₂/m-TiO₂/MAPbI₃/*spiro*-OMeTAD/Au) at different relative humidity (0, 30, 45 and 60 %), under dark conditions (see also Fig. S2)  (a) Cyclic voltammetry curves and (b) their respective hysteresis index defined by eqn. (1), calculated at 0.75 V. (c) Impedance plots with (d) the corresponding capacitance spectra, at 0.95 applied voltage. Solid lines correspond to fits using the equivalent circuit detailed in Fig 1c.

In Fig. 3a we show the cyclic voltammetry corresponding to the different %RH exposure. From Fig. 3a the evolution of hysteretic behaviour is clear: a switching from NH to IH is produced when moving from dry conditions to high humidity. For the best of our knowledge, there is not an equivalent definition of hysteresis index for dark conditions in literature, at least in the perovskite field. Then, an approach has been made following eq. 3 for dark hysteresis index (DHI):

$$DHI(V_R) = \frac{J_{RS}(V_R)}{J_{FS}(V_R)} - 1 \qquad (3)$$

were $J_{RS}$ and $J_{FS}$ are the current-density in the reverse and forward scan directions respectively, at the reference voltage $V_R$, established here in 0.75 V because is close to the maximum power point voltage under 1 sun. As it can be observed in Fig. 3b, a decrease in the DHI takes place with increasing %RH, i.e. from high positive values at 0% RH corresponding to NH until reaching a negative value at 60% RH corresponding to an IH.

In Fig. 3c (see Fig. S3b for zoom at low frequency region) it is shown the impedance response of the different cells measured at 0.95V. The impedance spectra corresponding to 0 and 30% RH do not present an arc below the Z' axis. However, when we increase the humidity above 45% RH an arc below the Z' axis appears, evidencing the signature of the negative capacitance (Fig. 3d). The quantification of these results applying the EC of Fig. 1b is shown in Table S1 in the SI. Here we focus on 45 and 60 % RH which are the conditions where the samples present negative capacitance, see Table 2.

**Table 2.** Extracted parameters from IS for the samples under moisture conditions showing negative capacitance.

| Cells | $R_{rec}$ ($\Omega \cdot cm^2$) | L (H·cm$^2$) | $R_L$ ($\Omega \cdot cm^2$) | $\tau_{kin}$=L/$R_L$ (s) |
|-------|------|------|------|------|
| 45% RH | 38 | 2.1 x 10$^4$ | 860 | **24** |
| 60% RH | 29 | 3.0 x 10$^3$ | 69 | **44** |

In both conditions, the kinetic constant present values in the range of ~$10^0$ - $10^2$ s, similar to the case of the bromide perovskite without Li, which agrees with large interactions between accumulated ions and the surface.[6] For these two moisture conditions, the $R_{rec}$ and $R_L$ parameters show clear differences. The lower values correspond to the higher %RH, which agrees with higher recombination rates. Therefore, as in the case of the bromide perovskite, the increase in negative capacitance yields to both, larger inverted hysteresis and an increment in recombination, suggesting both phenomena are related. The origin of this behaviour may be associated here to the chemical degradation of the TiO$_2$ and perovskite interfaces by moisture exposure, leading to a major density of defects, which can interact with the accumulated ions at the ESL/perovskite interface. This interaction might lead to the generation of an intermediate recombination state yielding to an additional recombination path, produced at the interface between the TiO$_2$ and the perovskite material. This new mechanism would yield a reduction in the overall recombination resistance.

To provide direct comparison between our bromide and iodide results we compare in Fig. 4 the results obtained for both series of data. In Fig.4a and 4b the kinetic times and the associated recombination resistances are represented for all samples respect to the applied bias. The kinetic times are in the same range for No-Li and moisture exposed devices, indicating that in these cases accumulated ions have strong interactions at the ESL/perovskite interface, dominating the hysteretic response of the CV. For Li devices, $\tau_{kin}$ is one order of magnitude lower, yielding to a behaviour dominated by ionic diffusion.

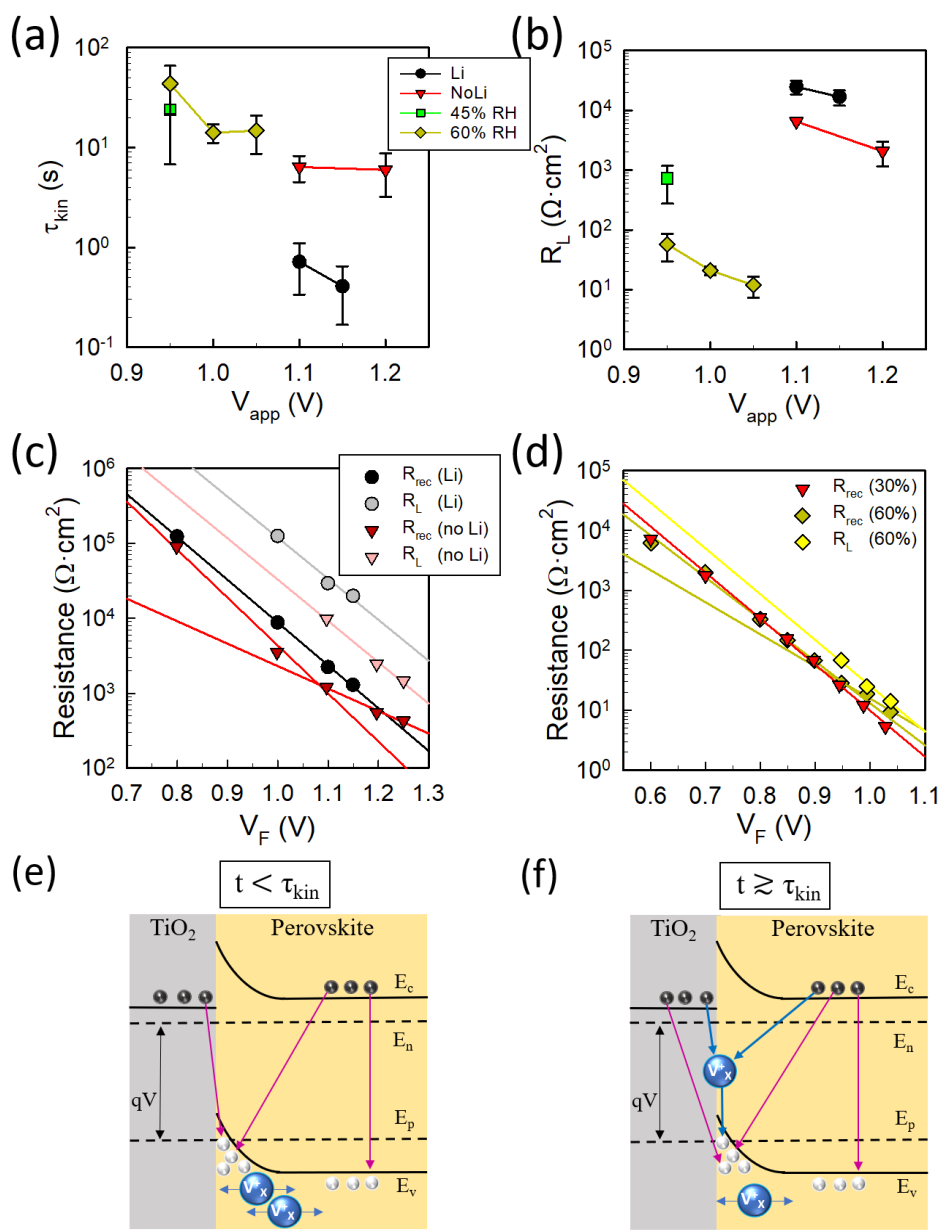

**Figure 4**. Kinetic relaxation times (a) and recombination values (b) versus applied bias corresponding to all samples analysed. It is shown how the kinetic relaxation time values and recombination resistances decrease with the applied voltage. Black circles and red triangles correspond to Li and No-Li samples, respectively; green squares and yellow diamonds to iodide samples under 45 %RH 65 %RH, respectively. (c) Evolution of $R_{rec}$ and $R_L$ with voltage corrected from the voltage drop at series resistance ($V_F = V - V_{drop}$). (d) The same for samples with 30 and 60% relative humidity. (e) Schematic diagram of the mechanisms occurring at times lower than $\tau_{kin}$. Black and white spheres represent electron and holes, respectively. Blue spheres represent ionic vacancies migrating towards the ESL/perovskite interface. (f) Shows the surface accumulation and binding of ions responsible for the large $\tau_{kin}$, favouring interfacial electronic recombination processes as well.

When we attend to the evolution of recombination resistance with voltage in Figs 4c and 4d, we observe that $R_{rec}$ is smaller than $R_L$ in all cases. This means that despite the overall recombination resistance (which is obtained as the parallel combination of $R_{rec}$ and $R_L$), is smaller than both, its behaviour is dominated by $R_{rec}$. Figs 4c and 4d show that for samples with low $\tau_{kin}$ or without negative capacitance, $R_{rec}$ presents a single exponential decrease with increasing voltage. The slope founded for $R_{rec}$ is similar to the one founded for $R_L$ (see table S2: $m_\beta$ and $m_\gamma$, ideality factors, defined in equations S6 and S7 in the SI). However, in samples showing negative capacitance, and precisely at the moment $R_L$ it starts to be observable, $R_{rec}$ diminishes its slope, increasing the ideality factors (see Table S2). This result suggests a change in the recombination mechanism that could move from being dominated by direct recombination of electrons with accumulated holes at the interface (Fig. 4e), to a recombination mediated by the interfacial states created by ionic vacancies (fig. 4f). A direct effect of the increasing in the ideality factor is a decrease of the J-V curve slope in the region of high voltages, leading to a lower FF (See eqs. S2 and S3).[41] In the Li samples this effect is minimal due to the fact that the final ideality factor is very similar. Contrary, for the samples with higher %RH, the flattening of the J-V curve is clearly appreciated in Fig. 3a and may be associated to the increased ideality factor yielding to the negative capacitance due to surface interactions.

As depicted in Fig. 4e when the kinetic relaxation time is small, the intrinsic ions can migrate towards the ESL/perovskite interface and accumulate, but not interact. This brings normal hysteresis and positive capacitive responses. However, when charges present large $\tau_{kin}$, the accumulated ions are allowed to interact with the surface and therefore, their movement to follow changes in the applied potential becomes delayed.

In summary, we have shown experimentally the intimate relationship between negative capacitance and IH. Employing the SPM, we get a kinetic relaxation time in the range of $10^0$ - $10^2$ s, which is associated to surface interactions with ions/vacancies at the ESL/perovskite interface. These interactions lead to a decrease in the overall recombination resistance, dominating low-frequency perovskite behaviour and yielding to negative capacitance response. Besides, the changes in recombination mechanisms and its values provoke a flattening of the cyclic voltammetry and the inverted hysteresis response. These combined effects yield to a decrease in *FF* and $V_{oc}$ (and thus in the performance of perovskite solar cells). We also have shown that this process is common in different type of perovskites (bromide and iodide) and under different treatments such as cation addition, which limits NC and thus IH, or humidity control, which acts the opposite way.

**Supporting Information**: Additional measurements, experimental part and further analysed recombination parameters.

**Acknowledgements**
This project has received funding from the European Union's Horizon 2020 research and innovation programme under the Marie Sklodowska-Curie grant agreement No 764787. Authors want to acknowledge Ministerio de Economía y Competitividad (MINECO) from Spain under project, ENE2017-85087-C3-1-R.